   \pdfoutput=1
   \documentclass[fleqn,usenatbib,usedcolumn]{mnras}
   \usepackage[british]{babel}             % British English hyphenation
   \usepackage{txfonts}                    % Reasonable fonts
   \let\la=\lesssim % for less than similar from txfonts, not \la from mnras.cls
   \usepackage[T1]{fontenc}                % Good font encoding
   \usepackage{graphicx}                   % Including figures
   \hypersetup{pdfauthor={G. Makiwa},
               pdftitle={Far-infrared/sub-millimetre properties of pre-stellar cores L1521E, L1521F and L1689B as revealed by the Herschel SPIRE instrument -- I. Central positions},
               pdfkeywords={stars: formation < Stars, (ISM:) dust, extinction < Interstellar Medium
		(ISM), Nebulae, infrared: ISM < Resolved and unresolved sources as a
		function of wavelength, submillimetre: ISM < Resolved and unresolved
		sources as a function of wavelength},
               bookmarksnumbered=true}
   \volume{{\rm in press}}
\usepackage{csquotes}

\def\HH{H$_2$}
\def\CO{CO}
\def\CS{CS}
\def\ceio{C$^{18}$O}
\def\ntwohplus{N$_2$H$^+$}
\def\hthcop{H$^{13}$CO$^+$}
\def\hcohp{HCOH$^+$}
\def\hcop{HCO$^+$}
\def\htco{H$_{2}$CO}
\def\thco{$^{13}$CO}
\def\thceio{$^{13}$C$^{18}$O}
\def\thsevco{$^{13}$C$^{17}$O}
\def\sevco{$^{17}$CO}

\def\tdust{T_\mathrm{d}}  % dust temperature
\def\mh{m_\mathrm{H}}  % mass of hydrogen atom
\def\nhh{N_\mathrm{H_2}}  % molecular hydrogen column density
\def\snhh{n_\mathrm{H_2}}  % molecular hydrogen number density
\def\pow#1#2{#1$\times$10$^{#2}$}
\def\nwerr#1#2{#1$\pm$#2}
\def\nwerrm#1#2{#1$\mp$#2}
\def\powpm#1#2#3{(#1$\pm$#2)$\times$10$^{#3}$}

\def\powmn#1{10^{#1}}
\def\micron{$\mu$m}
\def\lambdap{$\lambda_{\mathrm{peak}}$}
\def\rashort{$^{\textrm{h}\quad \textrm{m}\quad \textrm{s}}$}
\def\decshort{$^{\circ\quad \arcmin\quad \arcsec}$}
\def\sqcmpg{$\mathrm{cm^{2}~g^{-1}}$}  % W/m^2
\def\wsqmhz{$\mathrm{W~m^{-2}~Hz^{-1}}$}  % W/m^2/Hz^1
\def\wsqmhzst{$\mathrm{W~m^{-2}~Hz^{-1}~sr^{-1}}$}  % W/m^2/Hz^1/sr^1

\newcommand{\solarmass}{M$_{\odot}$}

\newcommand{\solarluminosity}{L$_{\odot}$}

\newcommand{\pscm}{cm$^{-2}$}
\newcommand{\pccm}{cm$^{-3}$}

\title[Far-infrared properties of pre-stellar cores]{Far-infrared/sub-millimetre properties of pre-stellar cores L1521E, L1521F and L1689B as revealed by the Herschel\thanks{{\it Herschel} is an ESA space observatory with science instruments provided by European-led Principal Investigator consortia and with important participation from NASA.} SPIRE instrument -- I. Central positions}

\author[G. Makiwa et al.]
{G. Makiwa,$^{1}$\thanks{E-mail:gibion.makiwa@uleth.ca} 
D. A. Naylor,$^{1}$ M. H. D. van der Wiel,$^{1,2}$ D. Ward-Thompson,$^{3}$ 
\newauthor
J. M. Kirk,$^{3}$ S. Eyres,$^{3}$ A. Abergel,$^{4}$ and M. K{\"o}hler$^{5}$ 
\\
$^{1}$Institute for Space Imaging Science, Department of Physics and Astronomy, University of Lethbridge, Canada\\
$^{2}$Centre for Star and Planet Formation, Niels Bohr Institute \& Natural History Museum of Denmark, University of Copenhagen,\\ \O ster Voldgade 5--7, DK-1350 \mbox{Copenhagen~K}, Denmark\\
$^{3}$School of Computing, Engineering and Physical Sciences, Jeremiah Horrocks Institute, University of Central Lancashire, UK\\
$^{4}$Institut d'Astrophysique Spatiale (IAS), Universit{\'e} Paris Sud \& CNRS, B{\^a}t. 121, Orsay 91405, France\\
$^{5}$School of Physics and Astronomy, Queen Mary University of London, Mile End Road, E14NS, London, UK
}

\date{Accepted 2016 February 22. Received 2016 February 22; in original form 2015 October 21}

\pubyear{2015}

\begin{document}
\label{firstpage}
\pagerange{\pageref{firstpage}--\pageref{lastpage}}
\maketitle

\begin{abstract}

Dust grains play a key role in the physics of star-forming regions, even though they constitute only $\sim$1 \% of the mass of the interstellar medium. The derivation of accurate dust parameters such as temperature ($\tdust$), emissivity spectral index ($\beta$) and column density requires broadband continuum observations at far-infrared wavelengths. We present \emph{Herschel}-SPIRE Fourier Transform Spectrometer (FTS) measurements of three starless cores: L1521E, L1521F and L1689B, covering wavelengths between 194 and 671 $\mu$m. This paper is the first to use our recently updated SPIRE-FTS intensity calibration, yielding a direct match with SPIRE photometer measurements of extended sources. In addition, we carefully assess the validity of calibration schemes depending on source extent and on the strength of background emission. The broadband far-infrared spectra for all three sources peak near 250 $\mu$m. Our observations therefore provide much tighter constraints on the spectral energy distribution (SED) shape than measurements that do not probe the SED peak. The spectra are fitted using modified blackbody functions, allowing both $\tdust$ and $\beta$ to vary as free parameters. This yields $\tdust$ of  \nwerr{9.8}{0.2} K, \nwerr{15.6}{0.5} K and \nwerr{10.9}{0.2} K and corresponding $\beta$ of \nwerrm{2.6}{0.9}, \nwerrm{0.8}{0.1} and \nwerrm{2.4}{0.8} for  L1521E, L1521F and L1689B respectively. The derived core masses are \nwerr{1.0}{0.1}, \nwerr{0.10}{0.01} and \nwerr{0.49}{0.05} $M_{\sun}$, respectively. The core mass/Jeans mass ratios for L1521E and L1689B exceed unity indicating that they are unstable to gravitational collapse, and thus pre-stellar cores. By comparison, the elevated temperature and gravitational stability of L1521F support previous arguments that this source is more evolved and likely a protostar.

\end{abstract}

\begin{keywords}
stars: formation -- dust, extinction -- infrared: ISM -- submillimetre.
\end{keywords}

%%%%%%%%%%%%%%%%%%%%%%%%%%%%%%%%%%%%%%%%%%%%%%%%%%

%%%%%%%%%%%%%%%%% BODY OF PAPER %%%%%%%%%%%%%%%%%%

\section{Introduction}\label{sec:starlesscoresintro}
The first stage in the formation of low-mass stars is considered to be the time when molecular clouds fragment into a number of cold ($T\leq$ 10 K) and dense ($\snhh \geq \powmn{5}$ \pccm) gravitationally bound cores \citep*[e.g.][]{andre2000prestellar}. These cores are usually referred to as starless or pre-stellar cores and were first identified in molecular line surveys carried out by Myers and co-workers \citep{myers1983dense, beichman1986candidate,benson1989survey}. Comparison of these cores with the IRAS point-source catalogue revealed that they were not associated with infrared sources, evidence that they represented the earliest stages of star formation. Observations of starless cores therefore provide an opportunity to study the initial conditions of protostellar collapse. 
\vspace{5cm} % I added this to avoid the following latex error which is caused by long references going from one page to another "pdfTeX error (ext4): \pdfendlink ended up in diff
%erent nesting level than \pdfstartlink.
%\AtBegShi@Output ...ipout \box \AtBeginShipoutBox                                                   \fi \fi 
%l.187    ./main_19January2015.tex:187:  ==> Fatal error occurred, no output PDF file produced!"

The cold and dense cores can be probed by studying their thermal (continuum) emission, which provides information about the dust properties (\citealp*[e.g.][]{andre1993submillimeter}; \citealp{ward1994submillimetre}), and by studying their line emission or absorption, which provides information on the chemistry in these regions (\citealp[e.g.][]{aikawa2001molecular}; \citealp{aikawa2005molecular}; \citealt*{pineda2008co}). 
Dense cores provide the environment necessary for the formation of molecules such as \HH, \CO\ and complex organic carbon bearing species \citep{herbst2009complex}. The homonuclear molecule \HH\ is difficult to observe in quiescent molecular clouds  and so the second most abundant molecule \CO\ is used as a molecular tracer. \CO\ is usually optically thick in molecular clouds and so its isotopologues (e.g. \thco, \ceio, \sevco, \thceio\ and \thsevco\ arranged in order of decreasing abundance) have been used to trace gas temperature and cloud mass in more opaque regions (\citealp[e.g.][]{bensch2001detection};\citealp{pineda2008co}). Infrared and millimetre observations of cold cores have provided evidence of freezing out of gaseous species onto dust grain surfaces \citep[e.g.][]{caselli1999co, tafalla2002systematic}. This freeze out results in a change in grain sizes, which affects the emissivity of the grains in ways that are currently not well understood (\citealp[e.g.][]{meny2007far}; \citealp*{sadavoy2010starless}).

In this paper we present far-infrared, broadband observations of three starless cores carried out using the Spectral and Photometric Imaging Receiver Array (SPIRE) instrument \citep{griffin2010herschel} on the Herschel Space Observatory \citep[hereafter referred to as \emph{Herschel,}][]{pilbratt2010herschel}. SPIRE consists of an imaging Fourier Transform Spectrometer (FTS) with two hexagonally packed bolometer arrays: SLW (447-990 GHz), and SSW (958-1546 GHz) and a Photometer with three hexagonally packed bolometer arrays:  PSW (250 \micron\ band), PMW (350 \micron\ band) and PLW (500 \micron\ band). Far-infrared observations from ground based telescopes are restricted to transmission ``windows'' in the atmosphere \citep[e.g.][]{holland1999scuba}. As a result, spectral energy distributions (SEDs) of starless cores have been constructed using flux densities obtained from a small number (3-5) of continuum maps (\citealp*[e.g.][]{kirk2005initial,kirk2007initial}; \citealp{schnee2007effect}). These maps are obtained using facility instruments on different telescopes having different beam sizes, each sensitive to structure on a different scale. Interpretation of SEDs obtained by combining data from different instruments taken under different atmospheric conditions is challenging. Fitting grey body functions to such SEDs using $\chi^{2}$ minimization techniques exposes the degeneracy between emissivity spectral index ($\beta$) and temperature ($\tdust$) \citep{juvela2012degeneracy, shetty2009effect, schnee2007effect}. \emph{Herschel} provided unfettered access to the far-infrared wavelength region, free from atmospheric absorption, and a stable operating platform. Since the peak of the SEDs for cold and dense cores falls within the range of the SPIRE FTS, analysis of the resulting spectra allows one to retrieve more robust values of $\tdust$, $\beta$ and column densities. 

This paper is the first in a series that will present results from our analysis of SPIRE FTS spectra of a number of star forming regions. In this paper we present the analysis of spectra corresponding to the brightness peaks of three starless cores, L1521E, L1521F and L1689B. The spatial extent and variation of dust temperature and emissivity spectral index will be the subject of a subsequent paper.

\section{Sources}\label{sec:sourceback}
\subsection{L1521E}\label{l1521Eback}
L1521E is located in the Taurus star-forming region at a distance of $d$=140 pc \citep{loinard2005multiepoch, torres2007vlba}. It is a triple-lobed pre-stellar core situated at the south-eastern end of the L1521 filament \citep{kirk2007initial}. L1521E was initially identified as a very young core by \citet*{hirota2002l1521e} who showed that the abundances of carbon-rich molecules in L1521E are higher than in other dark cloud cores and comparable to those in TMC-1. Studies carried out by \cite*{tafalla2004l1521e} have supported this claim by showing that L1521E has low \ntwohplus\ abundance and no \ceio\ depletion, implying that L1521E is less chemically processed and therefore recently contracted to its present density. 
\cite*{tafalla2004l1521e} estimate the age of L1521E to be $\le$ \pow{1.5}{5} yr. \citet{kirk2007initial} have studied L1521E and other starless cores by combining ground-based 24, 70 and 160 $\mu$m data from the Multiband Imaging Photometer (MIPS) on the \emph{Spitzer Space Telescope} \citep*[\emph{Spitzer},][]{werner2004spitzer}, 450 and 850 $\mu$m data from the Submillimetre Common-User Bolometer Array (SCUBA) on the James Clerk Maxwell Telescope (JCMT) and 90, 170 and 200 $\mu$m data from the imaging photopolarimeter (ISOPHOT) on the \emph{Infrared Space Telescope} (\emph{ISO}).
Despite the differences between these facility instruments, in particular their beam profiles, \citet{kirk2007initial} were able to derive core properties with a limited number of assumptions. 
Their modified blackbody fit to the L1521E SED with a fixed dust emissivity index ($\beta$ = 2) resulted in a dust temperature of \nwerr{8.1}{0.4} K. By using \hthcop\ observations of L1521E, \citet{hirota2002l1521e} derived a core radius of 0.031 pc, a mass of 2.4 \solarmass\ and a density of $\snhh$ = \pow{(1.3--5.6)}{5} \pccm\ for the brightest position. 

\subsection{L1521F}
L1521F (aka MC27; \citealt{codella1997four}; \citealt*{onishi1999very, lee2001survey}) is also located in the Taurus star-forming region at the same distance as L1521E. L1521F appears isolated with a strong central condensation as shown from the 160 $\mu$m \emph{Spitzer} map presented by \citet{kirk2007initial}. It is a dense starless core harbouring a low luminosity ($L$ = 0.05 \solarluminosity) object L1521F-IRS \citep{terebey2006spitzer, bourke2006spitzer}. Due to its low luminosity, L1521F did not appear in the IRAS catalogue \citep{beichman1986candidate, benson1989survey,codella1997four}, but was observed by \emph{Spitzer} \citep*{bourke2006spitzer}. L1521F-IRS together with L1544 \citep{crapsi2005probing} are the best known examples of evolved starless cores. \citet{kirk2005initial} obtained a dust temperature of 9 $\pm$ 2 K, a mass of \nwerr{0.4}{0.1} M$_{\sun}$ and a column density of $\nhh$ = \pow{1}{23} \pscm. The typical error in the derived $\nhh$ is $\pm$ 20-30 \%.

In their mapping surveys of Taurus, \citet{onishi1999very} argued that the high central density ($\sim$10$^{6}$ \pccm) and infall asymmetry seen in the \hcop(3-2) lines from L1521F indicate that the core is in its earliest stages of gravitational collapse with a free-fall timescale of 10$^{3}$ to 10$^{4}$ yr. Besides the high central density and infall asymmetry, L1521F also shows molecular depletion and enhanced deuterium fractionation \citep{crapsi2004observations, shinnaga2004physical}. \citet{crapsi2004observations} explain this by suggesting that L1521F is less chemically evolved and has a molecular hole of radius $<$ 2000 AU. \citet*{onishi1999very} have speculated that the spatially compact ($<$30\arcsec) line wings seen in \hcop (3-2) lines may be due to bipolar outflow emission similar to that seen from L1014-IRS by \citet*{huard2006deep}. A well-defined bipolar scattered light nebula seen at short wavelengths ($<$5 $\mu$m) by \citet*{bourke2006spitzer} also suggests the presence of molecular outflows in L1521F similar to those from low-mass protostars.

\subsection{L1689B}
L1689B is located in the Ophiuchus star-forming region at a distance of $d$=120 pc (\citealp{loinard2008preliminary}; \citealp*{lombardi2008hipparcos}). Millimeter and mid-infrared maps of L1689B have shown that it is an elongated core, sharp-edged in the north-south direction \citep{kirk2007initial}. \CS(2-1), \htco(2$_{12}$-1$_{11}$) and \hcohp(3-2) observations of L1689B have revealed line asymmetry typical of infall (\citealp*{lee1999survey}; \citealp{bacmann2000isocam}; \citealp{gregersen2000identify}). This suggests a supercritical core has formed and the object is entering the phase of dynamical contraction \citep{bacmann2000isocam, lee2001survey}. \citet*{jessop2001initial} compared their \ceio(2-1 and 3-2) observations and millimetre/submillimetre continuum observations of L1689B with results from a spherically symmetric radiative transfer model and concluded that freeze-out of CO is occurring towards the centre of L1689B.

\citet{kirk2007initial} and \citet*{roy2014reconstructing} fitted grey body functions to L1689B SEDs with a fixed dust emissivity index ($\beta$ = 2) and obtained dust temperatures of \nwerr{11}{2} K and 11.6 K respectively. \citet*{roy2014reconstructing} attempted to disentangle the effects of temperature variations along the line of sight by applying an inverse-Abel transform based technique and obtained a temperature of \nwerr{9.8}{0.5} K. 

The mass and column density of L1689B derived by \citet{kirk2005initial} are $M$ = \nwerr{0.4}{0.1} \solarmass\ and $\nhh$ = \pow{5}{22} \pscm\ respectively. Typical errors on $\nhh$ were reported to be $\pm$ 20--30 \%. \citet{roy2014reconstructing} derived a mass of 11 \solarmass.

\section{Observations and data reduction}\label{sec:starlesscoresobs}

\subsection{\emph{Herschel}-SPIRE and PACS photometer observations}\label{sec:photobs}
\begin{figure*}
\includegraphics[width=180mm]{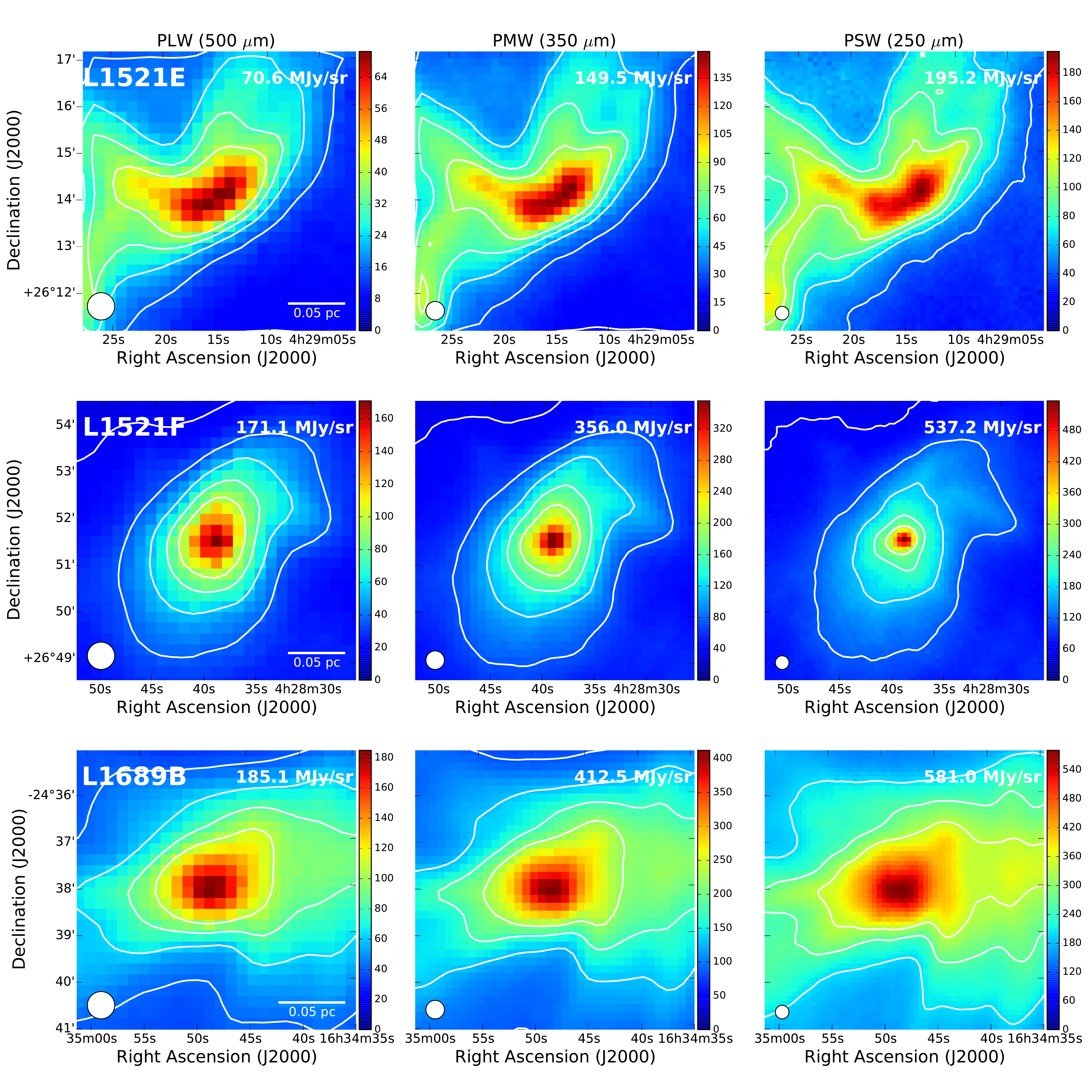}
\caption{The SPIRE photometer maps at 250 \micron\ (PSW), 350 \micron\ (PMW) and 500 \micron\ (PLW) for L1521E, L1521F and L1689B. Contours are shown starting from the brightest at 50, 40, 30, 20 and 10\% of the peak in each map. The peak flux density in MJy/sr for each map is shown at the top right corner. The contour at 10\% for L1521E and L1689B falls outside the maps. The circle at the bottom left corner of each map represents the photometer beam size at that wavelength. A scale bar corresponding to 0.05 pc is shown at the bottom right corner of each map.}
\label{fig:SPIREPhoto}
\end{figure*}

SPIRE and PACS photometer maps of the three starless cores: L1521E, L1521F and L1689B were obtained as part of the Gould Belt guaranteed time Key programme for the study of star formation \citep{andre2010filamentary}. This programme produced large area maps of the L1688 cloud in the rho Ophiuchus star forming region and the L1521 cloud in the Taurus star forming region. The L1521E and L1521F observations correspond to \mbox{OBSID} 1342202254 and that for L1689B to \mbox{OBSID} 1342205093. These observations were obtained using the SPIRE PACS parallel mode and processed using the Herschel Interactive Processing Environment (HIPE)\footnote{HIPE is a joint development by the Herschel Science Ground Segment Consortium, consisting of ESA, the NASA Herschel Science Center, and the HIFI, PACS and SPIRE consortia.} version 13.0 \citep{ott2010hipe}. Details for these observations are provided in Table~\ref{tab:obsdetailsall}. The data are publicly available through the Herschel Science Archive (HSA). Images of each core obtained at the three SPIRE photometer wavebands are shown in Figure \ref{fig:SPIREPhoto}.
\begin{table*}
\begin{center}
\caption{Observational details of starless cores in our ``Evolution of Interstellar dust'' key program: L1521E, L1521F and L1689B. These observations were obtained in sparse sampling mode. The first three columns show the target and its position (RA and Dec) on the sky. The last four columns show the instrument(s) used to observe the targets, the operational day (OD), the observation identification number (\mbox{OBSID}) and the duration of the observation. Since our cores occupy a tiny portion of the large ($>$1 square degree) photometer maps it is not instructive to include the duration for photometer observations.}
\begin{tabular}{llrlllr} 
\hline \hline
Target			&RA(J2000)	&Dec(J2000)	& Instrument   			&  OD&  \mbox{OBSID}  				&Duration\\ 
				&(\rashort)	&(\decshort)	&   					&  &	 					&[s] 	 \\ \hline
L1521E\_{on}		&04 29 13.9	&26 14 10.2	& SPIRE FTS			&  288&1342191211 				&1087 \\
L1521E\_{int}		&04 29 10.9	&26 16 37.1	& SPIRE FTS			&  288&1342191212 				&3583 \\
L1521E\_{off}		&04 29 10.6	&26 20 13.9	& SPIRE FTS			&  288&1342191213 				&3583 \\
%L1521			&04 20 23.3	&27 49 23.3	& SPIRE/PACS Photometers	&  OBSID&1342202254 				&-- \\
L1521F\_{on}		&04 28 39.5	&26 51 34.5	& SPIRE FTS			&  288&1342191208 				&1087 \\
L1521F\_{int}		&04 28 31.0	&26 53 22.4	& SPIRE FTS			&  288&1342191209 				&3583 \\
L1521F\_{off}		&04 28 22.4	&26 57 50.5	& SPIRE FTS			&  288&1342191210 				&3583 \\
L1521			&04 20 23.3	&27 49 23.3	& SPIRE/PACS Photometers	&  451&1342202254 				&-- \\
L1689B\_{on}		&16 34 48.3	&-24 38 02.9	& SPIRE FTS 			&  288&1342191221 				&1087 \\
L1689B\_{int}		&16 34 38.1	&-24 38 03.2	& SPIRE FTS			&  288&1342191222 				&3583 \\
L1689B\_{off}		&16 34 48.2	&-24 42 58.9	& SPIRE FTS			&  288&1342191223 				&3583 \\
L1688			&16 26 48.3	&-24 11 23.4	& SPIRE/PACS Photometers	&  499&1342205093,1342205094 	&-- \\
\hline \hline
\end{tabular}\label{tab:obsdetailsall}
\end{center}
\end{table*}%
The calibration uncertainties associated with SPIRE and PACS photometer maps is in the region of 5\% or less \citep{griffin2013flux, Balog2013}.

\subsection{\emph{Herschel}-SPIRE spectrometer observations}\label{sec:specobs}
\begin{figure*}
\begin{center}
\includegraphics[width=55mm]{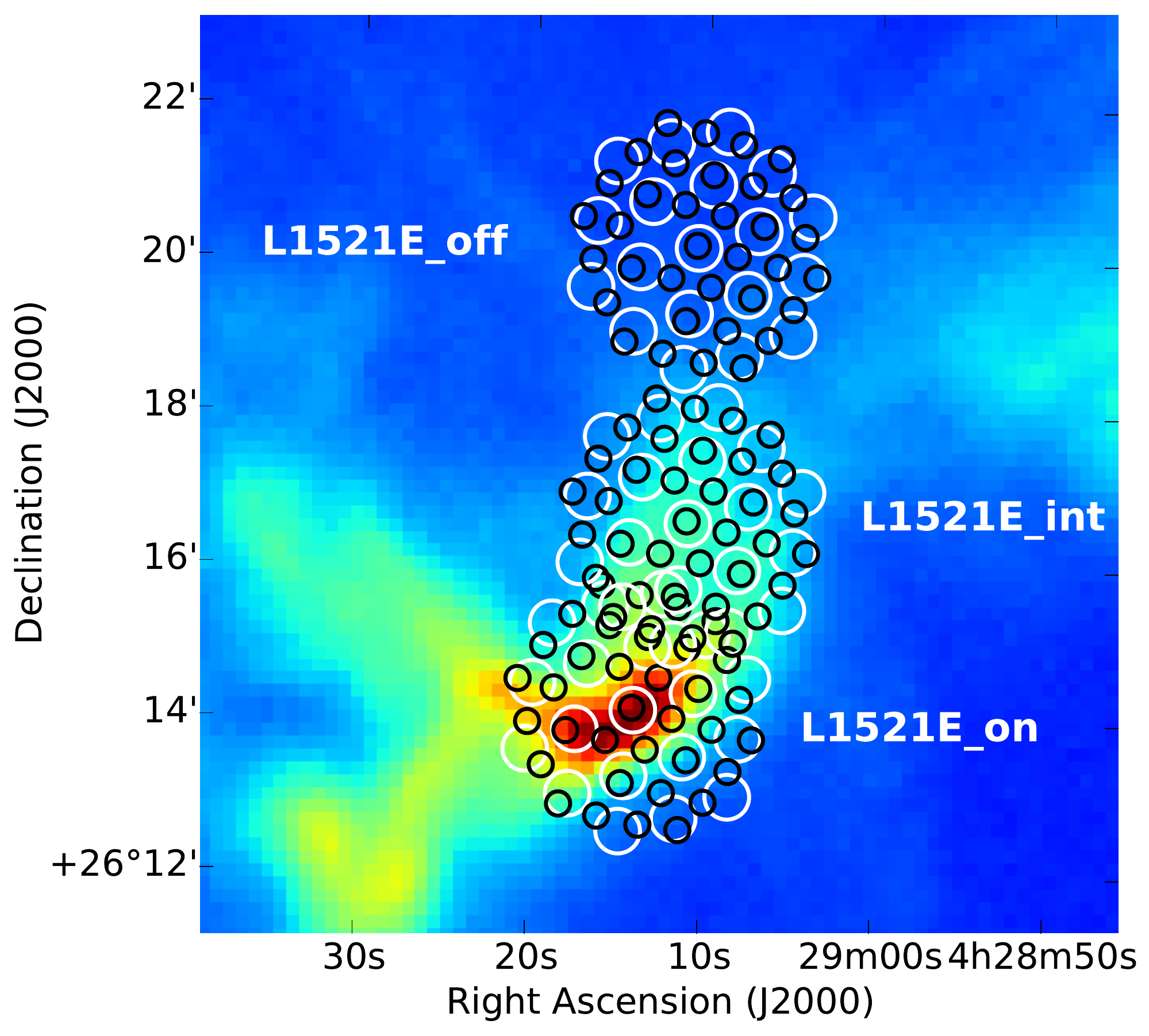}
\includegraphics[width=55mm]{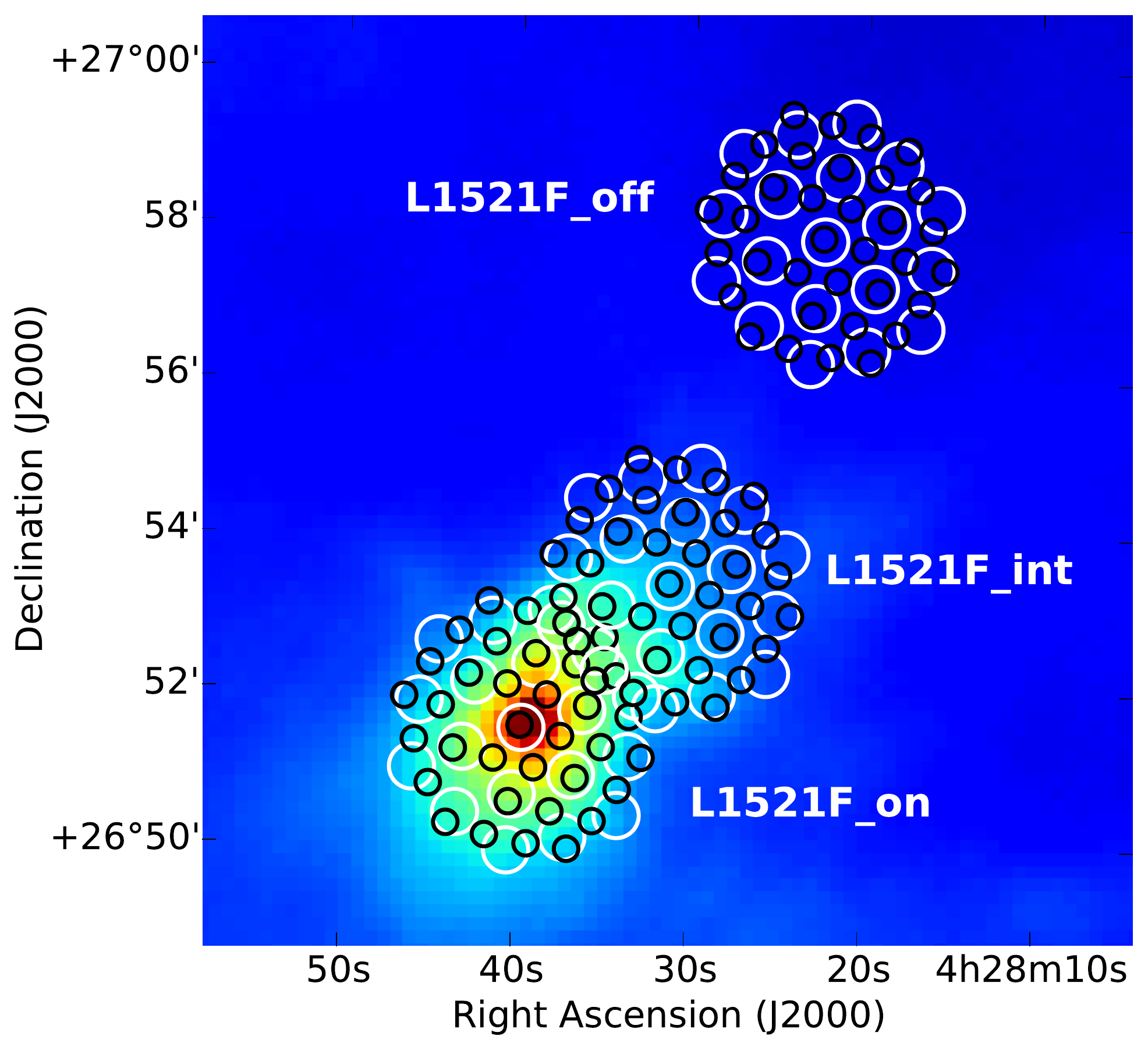}
\includegraphics[width=55mm]{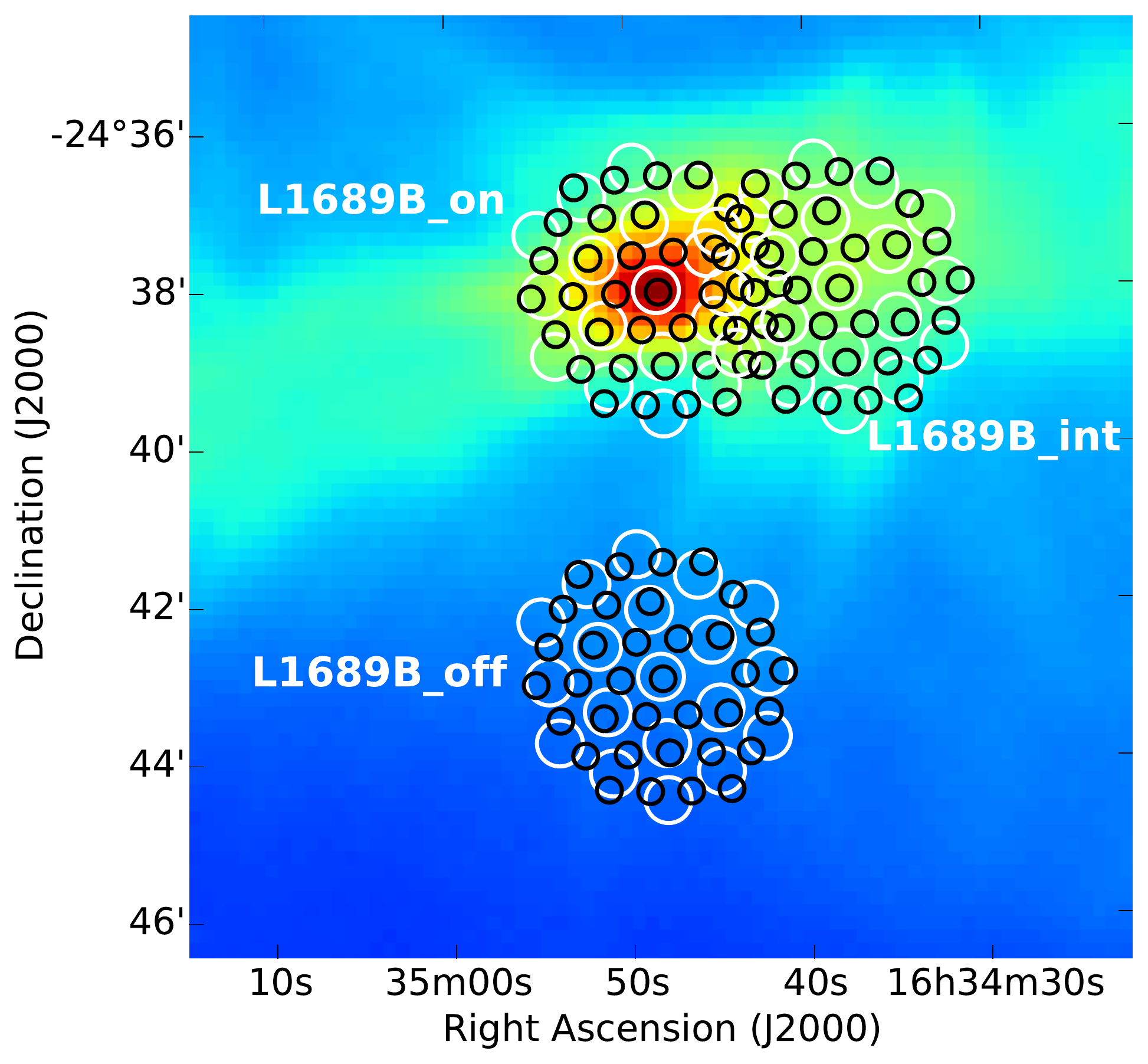}
\end{center}
\caption{SPIRE FTS footprints for three observations of L1521E (L1521E\_{on}, L1521E\_{int}, L1521E\_{off}), L1521F (L1521F\_{on}, L1521F\_{int}, L1521F\_{off}) and L1689B (L1689B\_{on}, L1689B\_{int}, L1689B\_{off}) over-plotted onto SPIRE Photometer PMW maps. The black and white circles represent the SSW and SLW detector arrays, respectively. The circle sizes correspond to a representative full width at half-maximum (FWHM) of the beam (19\arcsec\ for SSW detectors and 35\arcsec\ for SLW detectors).}
\label{fig:SPIREFootprints}
\end{figure*}

The starless cores (L1521E, L1521F and L1689B) were observed using the \emph{Herschel} SPIRE FTS as part of the \enquote{Evolution of Interstellar dust} key program under the ISM Specialist Astronomy Group number 4 \citep[SAG4,][]{abergel2010evolution}. A summary of the observation details is provided in Table~\ref{tab:obsdetailsall}. Three sparsely sampled observations were obtained for each core studied: \_{on} (on-source), \_{int} (intermediate or on the wings of the source) and \_{off} (off-source). The observations were taken in low spectral resolution mode, yielding a spectral resolving power ($\nu/\Delta \nu$) of $\sim$20 -- 60 across the frequency range 447 -- 1546 GHz (671 -- 194 \micron). Figure~\ref{fig:SPIREFootprints} shows the SPIRE FTS detector array footprints for these observations over-plotted onto SPIRE Photometer PMW maps. 

The SPIRE FTS data were reduced using HIPE version 14.0, which applies a major correction to all extended-source calibrated FTS spectra correcting for the far-field coupling efficiency of the FTS feed-horns (see \citealt{observers2014herschel} and Valtchanov et al. (in preparation) for details). Prior to HIPE version 14.0, flux densities of extended sources observed with the FTS were found to be significantly lower than those from the SPIRE photometer by factors of between 1.4 and 1.7. From the SPIRE FTS sparsely sampled observations, spectra were obtained for each of the detectors in the short wavelength (SSW) and long wavelength (SLW) arrays \citep{griffin2010herschel}. In this paper we discuss results from analyzing spectra for the central detectors (SLWC3 and SSWD4) for the SPIRE FTS sparse sampled observations pointed at the on-source positions of the three cores (Table~\ref{tab:obsdetailsall}).
The uncertainty in the derived intensity for sparsely sampled observations of reasonably bright and extended sources observed in the central detectors of the SPIRE FTS is 7 \% \citep{swinyard2014calibration}. 

\section{Results and discussion}\label{sec:starlesscoresresults}

\subsection{Morphology of far-infrared emission}\label{sec:scmorphology}
Figure \ref{fig:SPIREPhoto} shows the SPIRE photometer maps for each of the three starless cores; each map measures 6\arcmin$\times$6\arcmin\ and is centred on the source in question (Table~\ref{tab:obsdetailsall}). The SPIRE photometer beam at each wavelength is represented by a circle on the lower left corner of each map. On the lower right corner of each map is a linear scale in pc. Contours are shown at 50, 40, 30, 20 and 10\% of the peak flux density indicated and show that all three cores are extended with respect to the largest SPIRE photometer beam FWHM of 36.4\arcsec\ at 500 $\mu$m, which corresponds to 5096 AU = 0.025 pc at an assumed distance of 140 pc. For each source, the far-infrared images show that there is a similarity in the extent of the continuum emission, evidence that the 250, 350 and 500 $\mu$m maps are tracing the same dust component.

\subsection{Analysis of SEDs}\label{sec:starlesscoresanalysis}
In the limit of small optical depth, emission from dust particles can be described using a modified blackbody (grey body) function \citep[]{hildebrand1983determination}:
\begin{equation}\label{eqn:1}
F_{\nu}=\tau_{\nu} B_{\nu}(\tdust) \Omega= \mu \mh \nhh \kappa_{\nu} B_{\nu}(\tdust) \Omega \textmd{\qquad [Jy]},
\end{equation}
where $F_{\nu}$ is the flux density per reference beam in Jy (1 Jy=\pow{1}{-26} \wsqmhz) and $B_{\nu}(\tdust)$ is the Planck blackbody emission from dust at temperature $\tdust$:
\begin{equation}\label{eqn:2}
B_{\nu}(\tdust) = \frac{2h\nu^{3}}{c^{2}}\frac{1}{\exp{(h\nu/k\tdust)}-1} \textmd{   [\wsqmhzst]},
\end{equation}
$\tau_{\nu}$ is the optical depth, which can be expressed in terms of the emissivity of dust (specific opacity), $\kappa_{\nu}$ (in cm$^{2}$  g$^{-1}$), the density of the gas-dust mixture $\rho$, and the path length $L$ by \citep[e.g.][]{terebey2009far}
\begin{eqnarray}\label{eqn:2a}
\tau_{\nu}& = & \kappa_{\nu}\rho L\\
{} & = & \kappa_{\nu} \mu \mh\nhh\\
{} & = & \sigma_{\nu} \nhh.
\end{eqnarray} 
The emissivity of the dust at frequency $\nu$ is parametrized by \citet{beckwith1990survey} as:
\begin{equation}\label{eqn:3}
\kappa_{\nu} = \kappa_{\mathrm{1000}}\left(\frac{\nu}{1000 \textmd{ GHz}}\right)^{\beta} \textmd{ \qquad [\sqcmpg]},
\end{equation}
for a standard dust-to-gas ratio of 1:100. $\mu$ (= 2.8) is the mean molecular weight of interstellar material in a molecular cloud (with 71 \% molecular hydrogen gas, 27 \% helium and 2 \% metals) per hydrogen atom \citep[e.g.][]{kauffmann2008mambo}, $\mh$ is the mass of the hydrogen atom, $\nhh$ is the hydrogen column density, $\Omega$ is the beam solid angle subtended either by the reference beam (for spectra corrected using the semiExtendedCorrector tool (SECT) which will be discussed in Section \ref{sec:mismatch}) or by the adopted source extent (for extended calibrated spectra) and $\sigma_{\nu}$ is the opacity in units of cm$^{2}$ per hydrogen molecule. $h$ is the Planck constant, $k$ is the Boltzmann constant, $c$ is the speed of light, $\kappa_{1000}$ = 0.1 cm$^{2}$/g is the emissivity of dust grains at a frequency of 1000 GHz \citep{hildebrand1983determination}. We use the dust opacity law from \citet{hildebrand1983determination} to be consistent with other papers on dense cores \citep[e.g.][]{andre2010filamentary, sadavoy2013herschel}. However, recent results from dust models have shown that $\kappa_{1000}$ can vary by up to a factor of 2.0 \citep[e.g.][]{jones2013evolution}. 
$\beta$ is the dust emissivity index (1 $\leq \beta \leq 2$). Small values of $\beta$ ($\simeq$1) have been attributed to dust grains of mm size or larger \citep{beckwith1991particle} while values of $\beta$ between 1 and 2 have been attributed to changes in material composition of dust grains which result from grain growth \citep{ossenkopf1994dust, lis1998350, kohler2015dust}. The origin of values of $\beta\geq $ 2 is not yet well understood \citep{shirley2011observational}.

The total mass (gas and dust) contained within a circle of radius $D$/2 (where $D$ is the diameter in m) from the centre of the core, $M(r<D/2)$ can be obtained from Equation~\ref{eqn:1} \citep*{hildebrand1983determination} as 
\begin{equation}\label{eqn:4}
M = \frac{F_{\nu} d^{2}}{ \kappa_{\nu} B_{\nu}(\tdust)},
\end{equation}  
The number density $\snhh$ and hydrogen column density $\nhh$ are calculated as:
\begin{equation}\label{eqn:5}
\snhh = \frac{6}{\pi D^{3}}\frac{M}{\mu \mh},
\end{equation}
\begin{equation}\label{eqn:6}
\nhh = \frac{4}{\pi D^{2}}\frac{M}{\mu \mh},
\end{equation}
respectively. The diameter of each core is derived from the angular source sizes calculated using SECT (discussed in Section \ref{sec:mismatch}). 
In this way, fits of a grey body function to the far-infrared/submillimetre SEDs provide values for the dust mass and molecular hydrogen column density.

The derived masses for the three sources have been compared with their Jeans masses to determine if they are stable against gravitational collapse. 
The Jeans mass can be expressed as \citep[e.g.][]{sadavoy2010starless}:
\begin{equation}\label{eqn:7}
M_\mathrm{J} = 1.9 \left(\frac{\tdust}{10 \mathrm{ K}}\right) \left(\frac{R_\mathrm{J}}{0.07 \mathrm{ pc}}\right),
\end{equation}
where $R_\mathrm{J}$ is the Jeans radius which we assume to be equal to the core radius $(=D/2)$. Cores with $M/M_\mathrm{J} < 1.0$ are considered to be stable against gravitational collapse while cores with $M/M_\mathrm{J} > 1.0$ are unstable and will collapse to form protostars.

\subsection{Effects of the SPIRE FTS beam profile}\label{sec:datared}
The SPIRE instruments use conical feed horns to couple the telescope beam onto the individual detectors. The photometer uses single-mode feed horns whose beams are well described by Gaussian profiles with FWHM values of 18.1\arcsec, 24.9\arcsec\ and 36.4\arcsec\, for the PSW, PMW and PLW, respectively \citep{observers2014herschel}. To accommodate the broad spectral range of the two bands, the FTS feed-horns are by necessity multi-moded, which results in complex wavelength dependent beam profiles \citep{Makiwa2013beam}. \citet{Makiwa2013beam} used Hermite-Gaussian functions to approximate the SPIRE FTS beams and have provided coefficients that can be used to reconstruct the beam profiles at each wavelength in the SPIRE band.

The calibration of the SPIRE FTS is performed using solar system objects and other astronomical targets as outlined by \citet{swinyard2010flight,
swinyard2014calibration} and \citet{hopwood2015systematic}. The standard data reduction pipeline for the SPIRE FTS was designed to work for two extreme cases of objects: those whose spatial extent is smaller than the beam and those whose spatial extent is much larger than the beam \citep{observers2014herschel}. The calibration scheme for the former is referred to as point-source calibration and, for the latter, extended-source calibration. Neither of the calibration schemes work for sources whose spatial extent falls in-between the two extremes as is the case for L1521F. 

\begin{figure*}
\begin{center}
\includegraphics[width=0.65\textwidth]{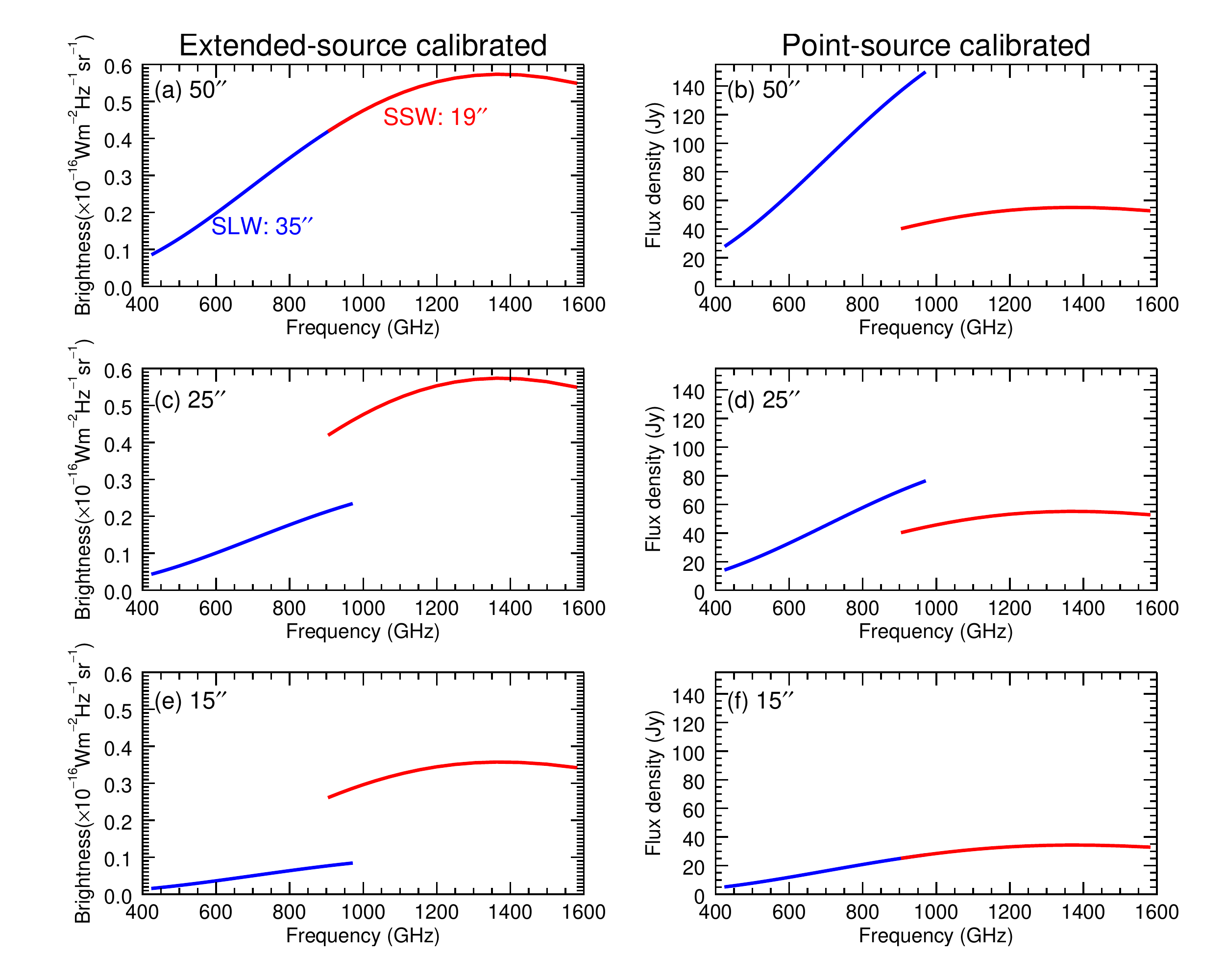}
\caption{The impact of assuming extended- and point-source calibration for a fictitious instrument similar to the SPIRE FTS when viewing a grey body of temperature 15 K and dust emissivity index of 1.45 and where the source size is 50\arcsec\ (top row), 25\arcsec\ (centre row) and 15\arcsec\ (bottom row), respectively. It is assumed that the instrument has a constant beam size of 19\arcsec\ in the SSW and 35\arcsec\ in the SLW bands, respectively. See text for a description of the figures.}
\label{fig:source_simulations}
\end{center}
\end{figure*}

A large gap and a difference in slope between the SSW and SLW spectra in the overlap region is an indicator that the applied calibration is not appropriate for the source being studied. This is a result of the fact that the beam diameter for the SLW detectors is almost twice as large as that for the SSW detectors in the overlap region. To illustrate this point, Figure \ref{fig:source_simulations} shows the impact of assuming extended- and point-source calibration for a fictitious instrument similar to the SPIRE FTS when viewing a grey body of temperature 15 K and dust emissivity index of 1.45 and where the source size is 50\arcsec\ (top row), 25\arcsec\ (centre row) and 15\arcsec\ (bottom row), respectively. To simplify the problem, it is assumed that the instrument has a constant beam size of 19\arcsec\ in the SSW and 35\arcsec\ in the SLW bands, respectively. The left column shows the brightness expected from the grey body when using extended-source calibration. The right column shows the spectra produced when using point-source calibration.
\begin{figure*}
\begin{center}
\includegraphics[width=\textwidth]{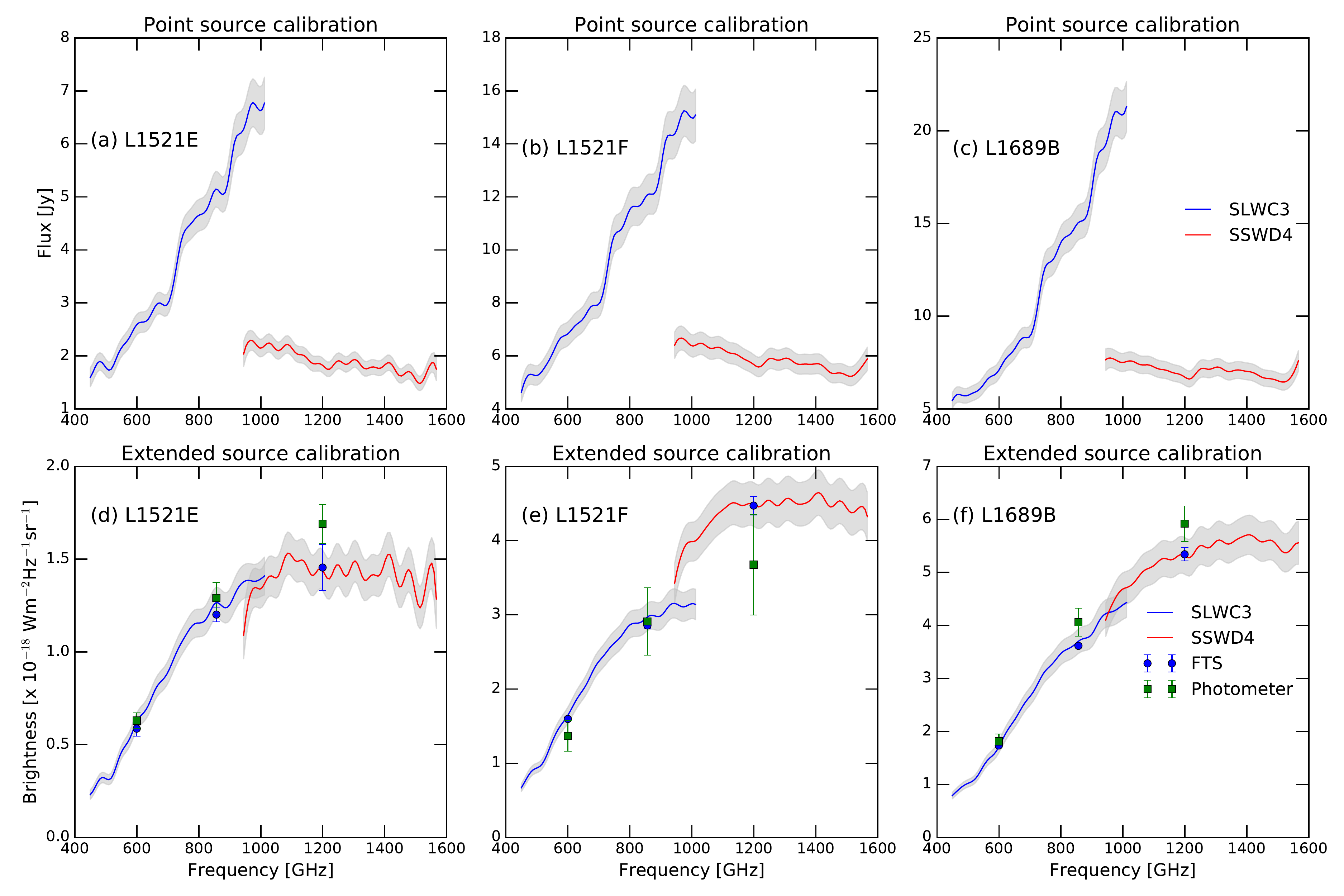}
\end{center}
\caption{SPIRE FTS spectra for L1521E (left column), L1521F (middle column) and L1689B (right column) after point-source (top row) and extended-source (bottom row) calibration. The SLWC3 spectrum is shown in blue and the SSWD4 spectrum in red. Grey bands represent intensity calibration uncertainties for the spectral observations. Flux densities from the SPIRE photometer are shown as green squares. The blue circles represent flux densities derived from the FTS spectra by integrating within the photometer bands.} 
\label{fig:Point_Ext_AllCores}
\end{figure*}
In the top row, the source size is 50\arcsec, which is larger than both the SSW and SLW beam sizes. Since in this case the source will appear extended to both beams, it is appropriate to use the extended source calibration which yields the continuous spectrum, seamless in the overlap region, as shown in the top left panel. However, if point source calibration was used it would lead to a large gap in the overlap region as shown in the top right figure. 

In the centre row, the source size is 25\arcsec, which is larger than the SSW beam size but smaller than that for SLW. Since in this case the source is extended to SSW and point-like to SLW, it is appropriate to use extended source calibration for SSW and point-source calibration for SLW. In the left figure, the spectrum for SSW (red) is correct, while the spectrum for SLW (blue) is not. In the right figure, the spectrum for SLW (blue) is correct while that for SSW (red) is not.

In the bottom row, a source size of 15\arcsec\ has been assumed. This is smaller than both the SSW and SLW beams. The source appears as a point source to both SSW and SLW and so applying extended-source calibration (left panel) is inappropriate for both bands. As is expected, applying point-source calibration (right panel) results in a continuous spectrum, again seamless in the overlap region.

Figure~\ref{fig:source_simulations} illustrates that important information on the source size is conveyed by studying the overlap region between the SSW and SLW bands. The gap is large when an observation of an extended source is point-source calibrated (top right panel). In this case the larger SLW beam measures more flux than the smaller SSW beam. The gap is smaller for a source size that is extended in the SSW band, but point-like in the SLW band (right panel in the centre row). Applying extended-source calibration to an observation of a source that is extended in the SSW band, but point-like in the SLW band (left panel in the centre row) results in a gap where the SLW spectrum is lower than the SSW spectrum. 

Point-source and extended source calibrated spectra for the centre detectors are shown in Figure~\ref{fig:Point_Ext_AllCores}. 
As discussed above, the large mismatch between point-source calibrated SLW and SSW spectra is an indication that these sources are not point-like. The mismatch is minimal for extended source calibrated spectra indicating that these sources are close to being fully extended. Of the three cores, extended source calibration works well for L1521E and L1689B while a significant difference exists in the overlap region for L1521F indicating that it must be approached differently. In fact, as will be shown in Section \ref{sec:mismatch}, L1521F is not as extended as the other two sources and extended-source calibration is therefore inappropriate. For comparison purposes, the bottom panels of Figure~\ref{fig:Point_Ext_AllCores} also show flux densities of the cores derived from extended-source calibrated SPIRE photometer maps, using circular apertures having the same size as the FTS beam, at the corresponding wavelengths. The photometer data shown represent the mean of the colour corrected flux densities of all map pixels in the circular apertures used. The error bars include both the statistical uncertainty from the spectra and the overall calibration uncertainties. 

\begin{figure*}
\begin{center}
\includegraphics[width=\textwidth]{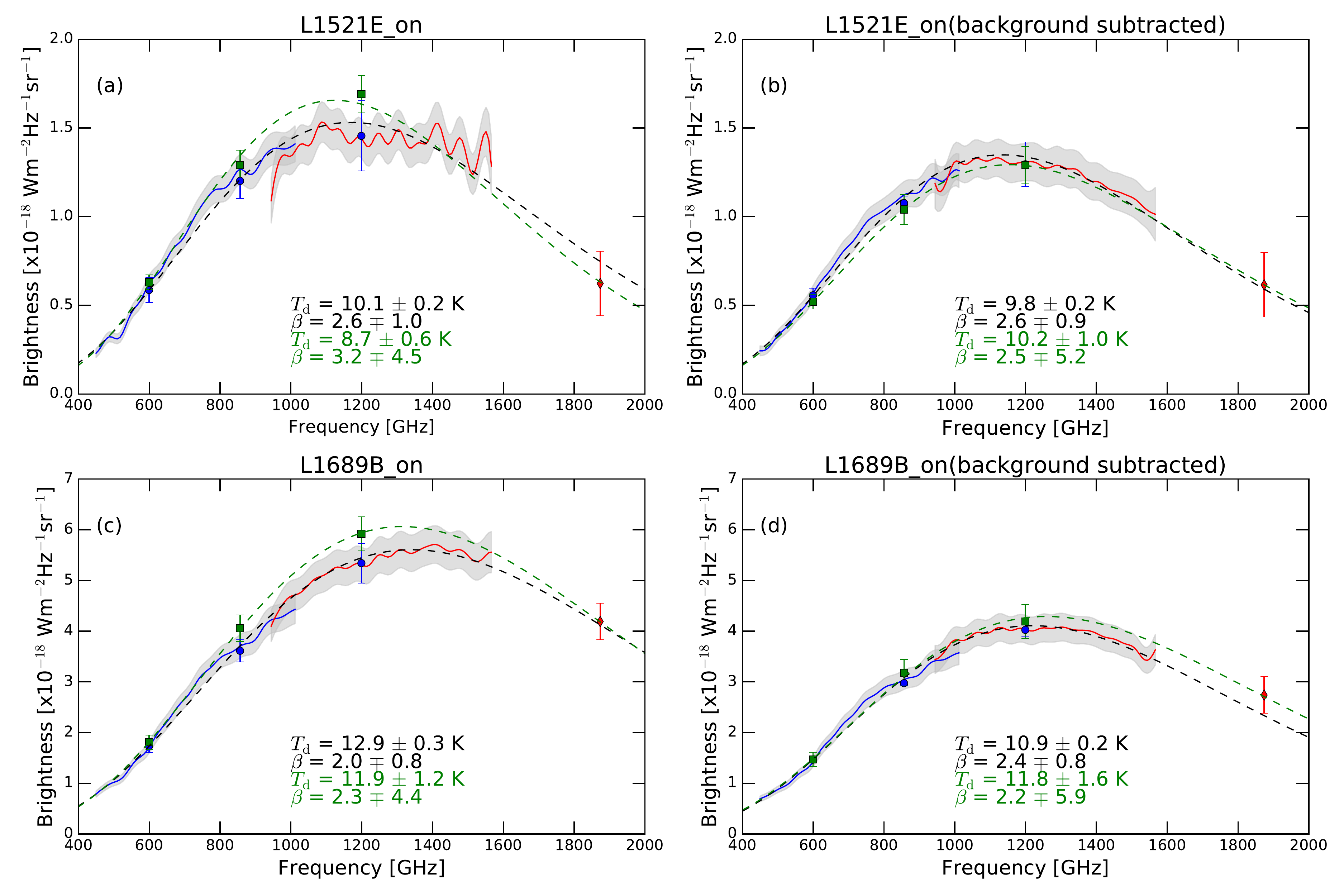}
\end{center}
\caption{SPIRE FTS extended-source calibrated spectra for L1521E (top row) and L1689B (bottom row) without background subtraction (left column) and with background subtraction (right column). The SLWC3 spectrum is shown in blue and the SSWD4 spectrum in red. Grey bands represent intensity calibration uncertainties for the spectral observations. Flux densities from the SPIRE and PACS photometers are shown as green squares and red diamonds respectively. The blue circles represent flux densities derived from the FTS spectra by integrating within the photometer bands. Grey body fits to the SEDs are shown as black and green dashed lines for the FTS and photometer data respectively.}
\label{fig:Ext_AllCores}
\end{figure*}

\begin{figure*}
\begin{center}
\includegraphics[width=\textwidth]{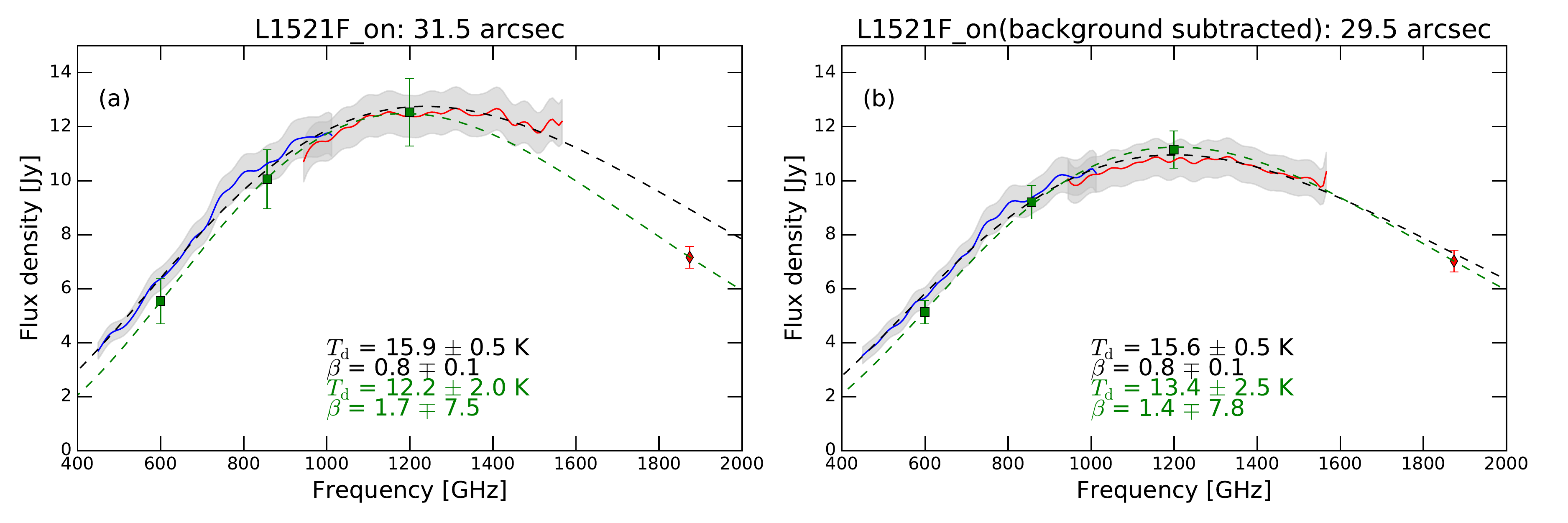}
\end{center}
\caption{SPIRE FTS point-source calibrated and SECT corrected spectra for L1521F without background subtraction (left panel) and with background subtraction (right panel). The SLWC3 spectrum is shown in blue and the SSWD4 spectrum in red. Grey bands represent the uncertainties as described in the text. The source sizes required to fix the gap in the overlap region for the uncorrected and background corrected spectra are 31.5\arcsec\ and 29.5\arcsec\ respectively.  Grey body fits to the SEDs are shown as black dashed lines. The derived dust temperatures from the uncorrected and background corrected spectra are 16.0 $\pm$ 0.5 K and 15.7 $\pm$ 0.5 K and the corresponding emissivity indices have the same value of 0.8 $\mp$ 0.1. Flux densities from the SPIRE and PACS photometers are shown as green squares and red diamonds respectively.}
\label{fig:Point_SECT_AllCores}
\end{figure*}

\subsection{Correcting for the mismatch between SLW and SSW spectra}\label{sec:mismatch}
Before the combined SLW and SSW spectra can be fitted with a grey body function it is essential that the gap and differences in slope in the overlap region that result from source-size issues be addressed. We have explored two methods of overcoming this challenge.

The first method is background subtraction. The background can be determined from a select number of off-axis detectors or from observations of the same region slightly offset from the target. The effect of emission from the local background in which cores would be found was recognized in preparing the ``Evolution of Interstellar dust'' key program under the ISM Specialist Astronomy Group \citep[SAG4,][]{abergel2010evolution}. As a result the proposal included three observations for each of the starless cores, two of which were pointed slightly away from the source as discussed in Section~\ref{sec:specobs} and Figure~\ref{fig:SPIREFootprints}. 

Figure \ref{fig:Ext_AllCores} shows the extended-source calibrated spectra for the on-source positions for L1521E (top left) and L1689B (bottom left) and the corresponding differences between extended-source calibrated spectra for the on-source and for the off-source positions (right column). In this paper only data from the central detectors (SLWC3 and SSWD4) from each footprint shown in Figure~\ref{fig:SPIREFootprints} is used. Spectra for the other detectors will be discussed in a subsequent paper focused on the spatial variation of dust parameters. Figure~\ref{fig:Ext_AllCores} shows that subtracting a background corrects for the differences in slope in the overlap region between SLW and SSW spectra. Fitting a grey body function to the resultant SED for L1689B results in a lower temperature than that from fitting the uncorrected SED. This is a clear indication that spectra measured by the SPIRE FTS beam consist of emission from both a hot extended foreground or background and a cold core. This is consistent with the interpretation that the centres of these self-gravitating cores are cold while the outer regions, which are heated by an external radiation field, are warmer. If the emission from the warmer dust in the outer layers is not removed from the measured data, the derived core masses will be biased towards lower values \citep{malinen2011accuracy}. Fitting to the L1521E SED -- a more extended core -- before and after background subtraction results in almost the same temperatures and dust emissivity indices implying that the background is a less important contributor to the measured flux. The relative importance of background flux can be seen by examining the spectra in Figure 5: for L1521E, the background contributes $\sim$10\% of the total brightness (Figure 5a vs. 5b), whereas the background contribution is $\sim$30\% for L1689B (Figure 5c vs. 5d). 

For comparison purposes, Figure~\ref{fig:Ext_AllCores} also shows flux densities of the cores derived from extended-source calibrated SPIRE and the Photodetector Array Camera and Spectrometer \citep[PACS, ][]{poglitsch2010photodetector} photometer maps using circular apertures having the same size as the FTS beam at the corresponding wavelengths for the on-source and off-source positions. The photometer data shown in Figure~\ref{fig:Ext_AllCores} represent the mean of the colour corrected flux densities of all map pixels in the circular apertures used. As mentioned previously, the error bars include contributions from the statistical analysis and calibration uncertainties. An equivalent background subtraction method as that used for the FTS spectra has also been applied to the photometer data and the resultant SEDs fitted with grey body functions. It can be seen that the brightnesses measured by the SPIRE and PACS photometers are consistent with those from the FTS, yielding consistent temperatures and dust emissivity indices especially for background subtracted SEDs. The derived parameters from fitting a grey body to the FTS spectra are based upon 45 independent spectral points whereas the fit to the photometer data is based upon four points. As a result FTS spectra provide a better constraint to grey body fitting. 

A similar exercise carried out on spectra from L1521F showed that background subtraction alone cannot correct for the gap in the overlap region between SLW and SSW spectra. Extended source calibration appears to be overcorrecting the spectra for this source and so L1521F is not as extended as the other two cores. 

A second method for correcting the gap in the overlap region between SLW and SSW spectra is to use the \enquote{semiExtendedCorrector} tool (SECT) as discussed by \citet{wu2013observing}. The tool corrects sparse sampled point-source calibrated spectra using the measured beam profile \citep{Makiwa2013beam} and an image of the source distribution. The spectrum is corrected to a default reference beam of 40\arcsec. In the case of L1521F, the spectra were corrected to a reference beam equal to the source size derived from SECT. SECT applies a correction for the forward coupling efficiency, but does not correct for effects such as the efficiency with which the reconstructed beam shape couples to the source, the deviation of the model from the true source distribution, and the response far from the axis \citep{observers2014herschel}. 
The tool is therefore limited to sources that are not extended. For this reason, SECT was only used to correct spectra for L1521F.

Figure~\ref{fig:Point_SECT_AllCores} shows the results from applying the SECT tool to point-source calibrated and background subtracted spectra for L1521F and fitting grey body functions to extract the dust temperature and emissivity index. Figure~\ref{fig:Point_SECT_AllCores} also shows colour corrected flux densities of the core derived from point-source calibrated SPIRE and PACS photometer maps. We used circular apertures derived by assuming that the measured map for L1521F is a convolution of a Gaussian source profile and the instrumental beam profile at the corresponding wavelengths for the on-source and off-source positions. The error bars are as described above.

The source sizes required to eliminate the gap in the overlap region using SECT for the background subtracted spectra are 61\arcsec, 29.5\arcsec\ and 39\arcsec\ which corresponds to 0.041, 0.019 and 0.023 pc for L1521E, L1521F and L1689B, respectively (the latest SECT updates do not allow determination of source sizes for sources larger than 60\arcsec\ since these are considered extended). These source sizes are in remarkable agreement with 0.031 pc derived by \citet{hirota2002l1521e} for L1521E and 0.015 and 0.024 pc derived by \citet{kirk2005initial} for L1521F and L1689B, respectively. The smallest source sizes for L1521E, L1521F and L1689B estimated from the photometer maps using the 50 \% contour are $\sim$80\arcsec, $\sim$35\arcsec\ and $\sim$120\arcsec\ respectively. It is clear that source sizes for L1521E, L1521F derived from the SPIRE FTS spectra and photometer maps are in agreement while those for L1689B are not. L1689B appears larger than the other two cores on the photometer maps most likely due to the more extended foreground/background. Performing background subtraction reveals a much smaller cold core. The derived source size for L1521F is slightly less than the smallest SLW beam size, but larger than that for the maximum SSW beam indicating that L1521F is semi-extended. The derived size for L1521E is a factor of $\sim$1.5 larger than the largest SLW beam indicating that L1521E is extended. The derived size for L1689B is slightly less than that for the largest SLW beam and therefore lies at the boundary between semi-extended and extended sources. 

For the reasons discussed in previous paragraphs, analysis for L1521E and L1689B is based on spectra corrected only for background emission (Figure~\ref{fig:Ext_AllCores} (b) and (d)) while for L1521F we use the spectrum corrected for background emission and for source extent by SECT (Figure~\ref{fig:Point_SECT_AllCores} (b)). Since the intermediate positions are closer to the on-source positions the possibility exists for the intermediate position to be contaminated by emission from the core itself. For this reason we have chosen the off-source position when performing the background subtraction. However, we find that using spectra from either the intermediate or off-source positions for the background subtraction produces consistent results for $\tdust$ and $\beta$. 

\subsection{Derived properties for each core}
Once corrected for effects of source spatial extent, the SPIRE FTS and SPIRE/PACS photometer SEDs were separately fitted using Equation~\ref{eqn:1} to derive $\tdust$ and $\beta$. In fitting the grey body function to our data we used the MPFIT function which employs the Levenberg-Marquardt technique to solve least square problems \citep{markwardt2009non}. Uncertainties in the input FTS spectra include calibration uncertainties and the standard deviation from averaging spectra for 55 repetitions (110 scans). Uncertainties in the input photometer data include calibration uncertainties and the standard deviation from averaging flux densities for all pixels within an aperture. The best grey body fits to the SEDs are shown in Figure \ref{fig:Ext_AllCores} (for L1521E and L1689B) and in Figure~\ref{fig:Point_SECT_AllCores} (for L1521F) and the results are presented in Table~\ref{tab:starlessresults}. This allows us to derive the core masses, densities and column densities of the cores (Section~\ref{sec:starlesscoresanalysis}). The error values in the derived quantities do not incorporate uncertainties in the value of $\kappa_{1000}$ (Equation \ref{eqn:3}).

\begin{table}
\begin{center}
\caption{Results obtained from fitting greybody functions to the SEDs for L1521E, L1521F and L1689B. The first two rows list angular sizes ($\theta_{\textmd{FWHM}}$) and corresponding linear scales of the cores while the third and fourth rows list the derived values of temperature and dust emissivity index. The fifth and sixth rows list the wavelengths and optical depths at which the peak of the SEDs occur. The remaining rows list the mass ($M$), density ($n_{\textmd{\HH}}$), column density ($N_{\textmd{\HH}}$), Jeans mass ($M_\mathrm{J}$) and $M/M_\mathrm{J}$ ratio of the cores.}
\begin{tabular}{lccc} 
\hline \hline \\ \vspace{2mm}
											&  L1521E   		&  L1521F  		&	 L1689B \\  
\hline
\vspace{2mm}
$\theta_{\textmd{FWHM}}$ [\arcsec]					& 61				& 29.5			& 39 			\\ \vspace{2mm}
$D$ [pc]									& 	0.041			& 0.020			& 0.023		\\ \vspace{2mm}
$\tdust$ [K]									& 9.8$\pm$0.2		& 15.6$\pm$0.5	& 10.9$\pm$0.2 \\ \vspace{2mm}
$\beta$										& 2.6$\mp$0.9   	& 0.8$\mp$0.1		& 2.4$\mp$0.8 \\ \vspace{2mm}
$\lambda_{\textmd{peak}}$ [\micron]					& 263	        		& 250			& 246 		\\ \vspace{2mm}
$\tau_{\lambda_{\textmd{peak}}}$ [$\times$10$^{-3}$]	& 3.8$\pm$0.4		& 2.5$\pm$0.3		& 8.5$\pm$0.8 	\\  \vspace{2mm}
$M$ [\solarmass]							  	& 1.0$\pm$0.1		& 0.10$\pm$0.01	& 0.49$\pm$0.05\\ \vspace{2mm}
$n_{\textmd{\HH}}$ [$\times$ 10$^{6}$ cm$^{-3}$]		& 0.047$\pm$0.005	& 0.044$\pm$0.005	& 0.15$\pm$0.01\\ \vspace{2mm}
$N_{\textmd{\HH}}$ [$\times$ 10$^{22}$ cm$^{-2}$]		& 0.79$\pm$0.09	& 0.37$\pm$0.04	& 1.4$\pm$0.1	\\ \vspace{2mm}
$M_\mathrm{J}$ [\solarmass]						& 0.55			& 0.43			& 0.34	\\

$M/M_\mathrm{J}$ 								& 1.7			 	& 0.2			& 1.5	\\
\hline \hline
\end{tabular}\label{tab:starlessresults}
\end{center}
\end{table}%

\subsubsection{L1521E}
The SED for L1521E from the SPIRE FTS data peaks at a wavelength of \lambdap = 263 \micron\ which is in the SPIRE band. Our derived dust temperature ($T_{d}$ = 9.8 $\pm$0.2) for L1521E is slightly larger than $\tdust$ = 8.1 $\pm$ 0.4 derived by \citet{kirk2007initial} by assuming a dust emissivity index of 2.
The value of $\beta$ derived in this work ($\beta$= 2.6 $\mp$ 0.9) is consistent with the often assumed value of $\beta$ = 2 based on studies of dielectric functions of silicate and graphite grains by \citet{draine1984optical}. The FWHM size of L1521E (0.04 pc) obtained from SECT is slightly larger than 0.031 pc reported by \citet{kirk2007initial}. 

The derived optical depth for L1521E and for the other two cores is $\ll 1$ as is expected in these cold and dense environments. The derived core mass is $M$ = \nwerr{1.0}{0.1} \solarmass\ which is less than 2.4 \solarmass\ derived by \citet{kirk2007initial}. The column density is $\nhh$ = \powpm{0.79}{0.09}{22} \pscm. Using our derived core diameter of 0.041 pc, the derived column density can be converted to a density of $\snhh$ = \powpm{4.7}{0.5}{4} \pccm\ which is 2.8$\sigma$ lower than the minimum in the range $\snhh$ = \pow{1.3--5.6}{5} \pccm\ derived by \citet{hirota2002l1521e}. The ratio of the core mass to the Jeans mass is 1.7 implying that L1521E is unstable to gravitational collapse. The conclusion drawn from this analysis is consistent with that presented in papers cited above and in Section~\ref{l1521Eback}, namely that L1521E is a very young core that has recently contracted to its present density. 

\subsubsection{L1521F}
Fitting to the SPIRE FTS SED for L1521F results in \lambdap\ = 250 \micron\, which is also in the SPIRE band and yields a temperature, $T_{d}$ of \nwerr{15.6}{0.5} K and a dust emissivity index of \nwerrm{0.8}{0.1}.
Our temperature is greater than 9 $\pm$ 2 K reported by \citet{kirk2007initial}, which again was obtained by assuming a dust emissivity index of 2.
The large differences in the derived temperatures is most likely due to the degeneracy between $\beta$ and $\tdust$, which is such that larger values of $\beta$ correspond to lower dust temperatures and vice versa. The higher temperature and lower core mass ($M$ = \nwerr{0.10}{0.01} \solarmass) that we obtain support the existence of a protostar in the L1521F core as reported by \citet{bourke2006spitzer}. The low value of $\beta$ could be explained by the change in the dust properties as the temperature of the protostar increases. It could also be due to temperature variations along the line of sight \citep[e.g.][]{juvela2012effect} and therefore complete modelling of the core, which is beyond the scope of this paper, is necessary to provide a definitive conclusion. To the best of our knowledge, this is the first time a far-infrared observation of a dense core has resulted in such a low value for $\beta$. Our low $M/M_\mathrm{J}$ ratio of 0.3 which is less than 1.0 is also an indication that the core is stable against gravitational collapse. The column density of the gas in the envelope is $\nhh$ = \powpm{0.37}{0.05}{22} \pscm. Using our derived core diameter of 0.02 pc, the column density of the envelope can be converted to a density of $\snhh$ = \powpm{4.4}{0.5}{4} \pccm.

\subsubsection{L1689B}
For L1689B, the SPIRE FTS SED peaks at \lambdap\ = 246 \micron\, which, again, is within the SPIRE band. We obtained a temperature of \nwerr{10.9}{0.2} K and a dust emissivity index of \nwerrm{2.4}{0.8}. Our derived temperatures are consistent with \nwerr{11}{2} K derived by \citet{kirk2005initial} which again was obtained by assuming a dust emissivity index of 2. We also derived a core mass of \nwerr{0.49}{0.05} \solarmass\ and a column density of \powpm{1.4}{0.1}{22} \pscm. Using our derived core diameter of 0.023 pc, the derived column density can be converted to a density of $\snhh$ = \powpm{1.5}{0.1}{5} \pccm.

The $M/M_\mathrm{J}$ ratio of 1.5 implies that L1689B is gravitationally unstable to collapse. In fact studies by \citet{bacmann2000isocam} and 
\citet{lee2001survey} have shown that L1689B is undergoing dynamical contraction.

\section{Conclusion}\label{sec:starlesscoresconcl}
The \emph{Herschel} Space Observatory provides a unique platform to study the far-infrared universe. Low spectral resolution SPIRE FTS spectra of three starless cores: L1521E, L1521F and L1689B have been analyzed to derive their dust properties. The calibration schema employed makes interpretation of SPIRE spectra challenging when sources are neither point-like nor extended. Since the three studied cores fall into this regime, there is a mismatch between the overlap regions of the SLW and SSW spectra. Two methods have been used to address this problem: background subtraction and the use of SECT. The resulting corrected spectra have been fitted with grey body functions. Previous studies of starless cores have been based on $\sim$3-5 points from photometer maps covering the SEDs sparsely. As a result, it has been common practice to assume $\beta$ = 2. The broadband SPIRE FTS spectra provide tighter constraints on the SED shape allowing us to have both the dust temperature and emissivity index to vary as free parameters. For comparison purposes, we have also fitted grey body functions to flux densities derived from SPIRE and PACS photometer maps.

Using our FTS spectra, we derive temperatures, $T_{d}$, of \nwerr{9.8}{0.2} K, \nwerr{15.6}{0.5} K and \nwerr{10.9}{0.2} K and dust emissivity indices, $\beta$, of \nwerrm{2.6}{0.9}, \nwerrm{0.8}{0.1}, \nwerrm{2.4}{0.8}, respectively for L1521E, L1521F and L1689B. Dust emissivity indices for L1521E and L1689B are consistent with $\beta$ = 2 usually assumed in modelling starless cores. We derive core masses of \nwerr{1.0}{0.1}, \nwerr{0.10}{0.01} and \nwerr{0.49}{0.05} \solarmass\ for L1521E, L1521F and L1689B, respectively. Since the peaks for the SEDs for all three cores fall within the SPIRE FTS band, we believe that our derived core masses are more accurate than those previously recorded in literature. Our derived core properties are robust and will help improve radiative transfer models of starless and pre-stellar cores. 

The dust in L1521F is significantly warmer that that for the other two cores suggesting that it is being heated by an internal radiation source. L1521F could now be a low-mass protostar. 

\section*{Acknowledgments}
\emph{Herschel} is an ESA space observatory with science instruments provided by European-led Principal Investigator consortia and with important participation from NASA. SPIRE has been developed by a consortium of institutes led by Cardiff Univ. UK and including Univ. Lethbridge (Canada); NAOC (China); CEA, LAM (France); IFSI, Univ. Padua (Italy); IAC (Spain); Stolkholm Observatory (Sweden); Imperial College London, RAL, UCL-MSSL, UKATC, Univ. Sussex (UK); Caltech, JPL, NHSC, Univ.Colorado (USA). This development has been supported by national funding agencies: CSA (Canada); NAOC (China); CEA, CNES, CNRS (France); ASI (Italy); MCINN (Spain); SNSB (Sweden); STFC (UK); and NASA (USA). 

The authors would like to thank Ivan Valtchanov and Trevor Fulton for their valued input in the processing of SPIRE FTS data. G.M., D.A.N. and M.H.D.vdW. acknowledge support from the CSA and NSERC and the latter also by the Lundbeck Foundation. Research at the Centre for Star and Planet Formation is funded by the Danish National Research Foundation and the University of Copenhagen's programme of excellence.

%%%%%%%%%%%%%%%%%%%%%%%%%%%%%%%%%%%%%%%%%%%%%%%%%%%



%\bibliographystyle{mnras}

%\input{../../aastex_macros}
%\bibliographystyle{mnras}
%\bibliography{../../PhD_Thesis_Citations_AuthorYear_S}

%%%%%%%%%%%%%%%%%%%%%%%%%%%%%%%%%%%%%%%%%%%%%%%%%%

%%%%%%%%%%%%%%%%% APPENDICES %%%%%%%%%%%%%%%%%%%%%

%\appendix
%%%%%%%%%%%%%%%%%%%%%%%%%%%%%%%%%%%%%%%%%%%%%%%%%%

% Don't change these lines
\bsp	% typesetting comment
\label{lastpage}
\end{document}